\documentclass[final,twocolumn]{elsarticle}
\usepackage{amssymb}
\usepackage{multicol}
\journal{Journal of Magnetism and Magnetic Materials}

\begin{document}
\def\bfs{\mathbf{S}}
\def\bfk{\mathbf{k}}
\def\bft{\mathbf{t}}
\def\bfe{\mathbf{e}}
\def\bfx{\mathbf{x}}
\def\tr{\mathrm{tr}}

\begin{frontmatter}

\title{Topological defects and critical phenomena in two-dimensional frustrated helimagnets}

\author{A.O. Sorokin}
\ead{aosorokin@gmail.com}

\address{Petersburg Nuclear Physics Institute, NRC Kurchatov Institute, 188300 Orlova Roscha, Gatchina, Russia}

\begin{abstract}
Using a simple model of a frustrated helimagnet, the critical behavior is numerically investigated for planar or isotropic spins, and for cases of one or two chiral order parameters. The helical structure in this model arises from the competition between exchange interactions of spins of the first two range orders in one direction (in both directions) of a square lattice. The main result is that the critical and temperature behavior is primarily determined by topological defects that are present in all cases.
In the case of planar spins, vortices, fractional vortices and domain walls are present in the system. Their interaction leads to the appearance of the phase of a chiral spin liquid, or induces a single first-order transition, and in the vicinity of the Lifshitz point vortices lead to a reentrant phase transition to the phase with a collinear quasi-long-range order. When transitions in the chiral and continuous order parameters are separated in temperature, they are of the 2D Ising and Kosterlitz-Thouless types correspondingly. In the case of isotropic spins, so-called $\mathbb{Z}_2$ vortices are present. They do not lead to the appearance of a phase with long-range or quasi-long-range order in the case of one chiral order parameter. However, their interaction leads to a sharp change in the temperature dependence of the correlation length (crossover). In the case of two chiral parameters, there are long-range chiral order of the Ising type (chiral spin liquid) and domain walls. However, as a result of the interaction of vortices and walls, the crossover and chiral transition occur at the same temperature as a first-order transition.

\end{abstract}


\end{frontmatter}

\section{Introduction}

Magnetic helix is a simple example of a long-periodic modulated magnetic structure (see \cite{Isyumov84} for a review). Depending on its origin mainly determining a symmetry of a structure, a helix can be frustrated or non-frustrated. A helix of the first type arises in antiferromagnets due to a competition of exchange interactions between spins of several first range orders or due to a geometry of a lattice \cite{Villain59,Yoshimory59,Kaplan59}. A non-frustrated helix arises due to the Dzyaloshinskii-Moriya interaction \cite{Bak80,Nakanishi80} and is not considered in the current study.

For the theory of phase transitions, the helices are interesting in that they are an example of a system with a complex structure of an order parameter, differing from the usual $O(N)$ model and magnets with a collinear spin ordering described by this model. Recently, we have discussed the critical behavior of frustrated helimagnets in three dimensions \cite{Sorokin14} and have found that a phase transition is of first order induced by fluctuations for all values of the number of the chiral order parameters ($K=1,\,2,\,3$) and the number of spin components ($N=2,\,3$).

In two dimensions, the critical behavior in frustrated helimagnets is more diverse and strictly depends on the values of $N$ and $K$. Generally speaking, the two-dimensional criticality has a specificity caused by the Mermin-Wagner theorem forbidding spontaneous breaking of a continuous symmetry at non-zero temperature in systems with local interactions. So, at first sight, one may expect that the critical behavior corresponds only to discrete (chiral) order parameters and Ising-type phase transitions. In fact, a situation is more complicated.

Frustrated XY helimagnets (with planar spins, $N=2$) belong to the class of systems characterized by a combination of continuous and discrete degeneracies of the ground state (see \cite{Korshunov06} for a review). Thermal and critical properties of the system are associated with the existence at low temperature of a quasi-long-range order with a continuous $U(1)\equiv SO(2)$ order parameter and a long-range order with discrete $\mathbb{Z}_2$ order parameters. For the case $K=1$, with temperature increasing, the ordering breaks in two steps \cite{Sorokin12}, at least far away from the Lifshitz point separating collinear (commensurate) and helical (incommensurate) ordered phases. At first, a Berezinskii-Kosterlitz-Thouless (BKT) transition related to a continuous order parameter occurs, and then one observes a chiral transition related to a discrete order parameter and belonging to the universality class of the two-dimensional Ising model. The phase between two transitions can be characterized as a chiral spin liquid (see \cite{Starykh15} for a review). For the case $K=2$, BKT and both chiral transitions occur at the same temperature as a first-order transition \cite{Sorokin12-2}.

In the case $N=3$ and $K=1$, there is no phase transition at non-zero temperature. Nevertheless, in such systems, one observes an explicit crossover \cite{Azaria01} at finite temperature separating the low-temperature $O(4)$-sigma-model-like behavior and high-temperature one. And as we have shown recently in \cite{Sorokin17}, there is a single first-order transition occurs in the case $N=3$, $K=2$.

In this work, we continue to investigate all four cases of frustrated helimagnets using the simple lattice model and Monte Carlo simulations. In particular, we extend the previous results \cite{Sorokin12,Sorokin12-2,Sorokin17} and show completed phase diagrams, including the vicinity of the Lifshitz point and the reentrant transition from the disordered phase to one with the collinear quasi-long-range spin ordering, discussed in \cite{Okwamoto84,Pokrovsky14,Dimitrova14}. Brief historical reviews of previous investigations of frustrated helimagnets and related systems are placed in corresponding subsections of sect. \ref{sect3}.

It should be noted that in all cases thermal and critical behaviors of helimagnets are determined by topological defects. Vortices and domain walls are presented in the $N=2$ case, so-called $\mathbb{Z}_2$ vortices are presented in $N=3$, $K=1$, and $\mathbb{Z}_2$ vortices and domain walls determine the critical behavior in the case $N=3$, $K=2$ (see table \ref{tab}). In fact, besides vortices and domain walls, the $N=2$ case contains fractional vortices in the spectrum of topological excitations, which are kinks of domain walls, firstly discussed in the sophisticated model of a helimagnet \cite{Villain91,Uimin94}. One expect that fractional vortices and a non-perturbative interaction between vortices and walls may significantly change the critical behavior. At least, they are crucial for a sequence of phase transitions. In this work, monitoring properties of topological defects, we show that the critical behavior in the $K=1$ case at the chiral transition point remains universal falling in the two-dimensional Ising class. Moreover, we find that the critical value of the domain wall density is the same as in the pure two-dimensional Ising model \cite{Sorokin18}, but the critical properties of vortices are consistent with the Ising-XY model \cite{Sorokin18-2} instead of the pure $O(2)$ model with a BKT transition.

{\begin{table}[t]
\center
\caption{Possible order parameter spaces and types of the critical behavior in two-dimensional frustrated helimagnets. Topological defects: $v$ --- vortices, $w$ --- domain walls, and $\mathbb{Z}_2$ marks $\mathbb{Z}_2$ vortices.}
\small
\label{tab}
\begin{tabular}{cc|c|c|c}
\hline
\hline
$N$, & $K$ & $G/H$ & Transitions & Defects\\
\hline
$2$, & $1$ & $\mathbb{Z}_2\otimes SO(2)$ & BKT + Ising & $v+w$\\
$2$, & $2$ & $\mathbb{Z}_2\otimes\mathbb{Z}_2\otimes SO(2)$ & I order & $v+2w$\\
$3$, & $1$ & $SO(3)$ & Crossover &  $\mathbb{Z}_2$\\
$3$, & $2$ & $\mathbb{Z}_2\otimes SO(3)$ & I order &  $\mathbb{Z}_2+w$\\
\hline
\hline
\end{tabular}
\end{table}}

\section{Model and methods}
\label{sect2}

The Hamiltonian of the proposed model is
\begin{equation}
    H=\sum_{{\bfx},\mu}\left(J_1\mathbf{S}_{\bfx}\bfs_{\bfx+\bfe_\mu}+J_3^\mu\bfs_{\bfx}\bfs_{\bfx+2\bfe_\mu}\right),
    \label{model}
\end{equation}
where $\bfs_{\bfx}$ is classical $N$-component vector in cite $\bfx$ of a square lattice, $J>0$, and $\mu=1,\ldots K$ enumerates two directions of a lattice with the unit vectors $\bfe_1$ and $\bfe_2$. When $K=1$ and $J_3^1>J_1/4$, the ground state is a helix with a wave vector $\mathbf{q}_0=(q_0^1,0)$, $\cos q_0^1=-J_1/4J_2$, when $K=2$, the system at zero temperature falls in the helicoidal phase with two chiral order parameters with corresponding wave vector $\mathbf{q}_0=(q_0^1,q_0^2)$, where $\cos q_0^\mu=-J_1/4J_3^\mu$, and $\mu=1,\,2$. The spin ordering is planar and can be described by two orthogonal $N$-vectors. For $N=2$ spins, configurations $(\pm q_0^1,0)$ do not pass into each other via global spin rotations $SO(2)$. This adds to the order parameter space $G/H\sim SO(2)$ the extra discrete factor $\mathbb{Z}_2$ for $K=1$ and $\mathbb{Z}_2\otimes \mathbb{Z}_2$ for $K=2$. But for $N=3$ spins, configurations $(\pm q_0^1,0)$ are equivalent, so the discrete factor appears only for $K=2$. The order parameter spaces for all four cases are also shown in table \ref{tab}.

A numerical investigations of helical structures have a specificity associated with the helix incommensurability and its thermal properties. Even if one consider the relations of exchange interactions values corresponding to a multiple to a lattice constant helix pitch, a thermal renormalization of coupling constants, acting unequally to exchanges of different range orders, makes a helix incommensurate at finite temperature. A temperature increase of a helix pitch in a system with periodic boundary condition brings an additional tension at non-zero temperature due to the quantization of a helix pitch. In other words, a lattice size is multiple to a pitch, and even if one chooses a commensurate lattice size and the optimal helix pitch $q_0$ at zero temperature, this value $q_0$ becomes not optimal with the temperature increasing and does not correspond of the global free energy minimum. So, the most correct way is to use fluctuating boundary conditions \cite{Saslow92,Saslow98}. Moreover, this is necessary in a vicinity of the Lifshitz point $J_3=J_1/4$, where a helix pitch and its thermal expansion are sufficient large. Unfortunately, the choice of fluctuating boundary conditions leads to additional errors that are significant for small lattice sizes.

Really, one can use periodic boundary conditions for simulations away from the Lifshitz point on small lattices. Using the numerical results \cite{Sorokin12}, we can estimate a upper limit of a lattice size under which the commensurate helix pitch remains optimal even at the transition temperature. Lets consider the helicity modulus. It is defined as the increase in the free energy density $F$ due to a small twist $\Delta_\mu$ across the system in the direction $\bfe_\mu$
\begin{equation}
    \Upsilon_{\mu}=\left.\frac{\partial^2 F}{\partial{\Delta_\mu}^2}\right|_{\Delta_\mu=0}.
    \label{Y=dF po dd}
\end{equation}
For the model (\ref{model}), one finds
\begin{equation}
    \Upsilon_1 = \left<E''_1\right>-\frac{L^2}{T}\left<(E'_1)^2\right>+\frac{L^2}{T}\left<E'_1\right>^2,
    \label{Yy}
\end{equation}
where $T$ is temperature,
$$
    E'_1 =-L^{-2}\sum_\mathbf{x}\left(J_1\sin\tilde\Delta_1+2J_3^1\sin2\tilde\Delta_1\right),
$$
$$
    E''_1 =-L^{-2}\sum_\mathbf{x}\left(J_1\cos\tilde\Delta_1 +4J_3^1\cos2\tilde\Delta_1\right),
$$
with $\tilde\Delta=q_0+\Delta_1$. Since $q_0$ does not correspond to the free energy minimum, one finds $\left<E'_1\right>\neq0$ at $\Delta_1=0$. Using the numerical value of $\left<E'_1\right>$, we can estimate the twist corresponding to the minimum of the free energy:
\begin{equation}
    \Delta_1(T)=q(T)-q_0\approx\frac{\left<E'_1(T)\right>}{\left<E''_1(0)\right>}.
\end{equation}
Considering $J=1$, $J_3=0.5$, one obtains $\left<E''_1(0)\right>=\frac32$ and finds in fig.3 of ref. \cite{Sorokin12} that $\left<E'_1(T\approx T_\mathrm{v})\right>\approx0.03$, where $T_\mathrm{v}$ is the BKT transition temperature. Thus, $\Delta_1\approx0.02$. The step of the helix pitch quantization is $\frac{2\pi}{L}$, with $L$ is a lattice size. While $\Delta_1<\frac{\pi}{L}$, the commensurate helix pitch $q_0$ remains the closest to the free energy minimum. Therefore, one can estimate a upper limit of a lattice size as $L\lesssim150$.

In this paper, we investigate model (\ref{model}) using Monte Carlo simulations based on the over-relaxed algorithm \cite{Brown87,Creutz87}. We consider $L=24,\,36,\,48,\,60,\,90$ and $120$. For $J_3/J_1>0.3$, we use periodic boundary conditions, for the rest cases, we use fluctuating boundary conditions. Thermalization is performed within $6\cdot10^5$ Monte Carlo steps per spin, and calculation of averages within $6\cdot10^6$ steps. Details of transition temperature estimation methods for all types of transitions have been discussed in ref.\cite{Sorokin18}. For an Ising-like transition, we use Binder cumulant crossing method \cite{Binder81}. For a BKT-like transition, the Weber-Minnhagen finite-size scaling method \cite{Weber88} has been used. A temperature of a crossover induced by $\mathbb{Z}_2$ is estimated as a value at which the helicity modulus dependence on a lattice size $L$ ceases to correspond to the $O(4)$ sigma model \cite{Azaria92}.

As long as we consider commensurate helices, for the $N=2$ case the $SO(2)$ order parameter can be introduced as
\begin{equation}
    \mathbf{m}_i=\frac{n_{sl}}{L^2}\sum_{\mathbf{x}_i}\mathbf{S}_{\mathbf{x}_i},\quad
    \overline{m}=\sqrt{\frac1{n_{sl}}\sum\limits_{i}\left<\mathbf{m}_i^2\right>},
    \label{m}
\end{equation}
where index $i$ enumerates $n_{sl}$ sublattices. In a vicinity of the Lifshitz point, to study a possible quasi-long-range order collinear phase, we also monitor the usual (anti)ferromagnetic magnetization. The chiral (discrete) order parameters are
\begin{equation}
    k_\mu=\frac{1}{L^2\sin\theta_0}\sum\limits_\mathbf{x} \sin(\varphi_\bfx-\varphi_{\bfx+\bfe_\mu}),
    \quad \overline{k}=\sqrt{\left<k^2\right>},
    \label{kiral2}
\end{equation}
with $\bfs_\bfx=(\cos\varphi_\bfx,\sin\varphi_\bfx)$.

The $SO(3)$ order parameter $\Phi_\bfx=(\bfs_\bfx,\bfk_\bfx)$ is a $3\times 2$ matrix composed of two orthogonal unit $3$-vectors $\bfs_\bfx$ and $\bfk_\bfx$, where
\begin{equation}
    \bfk_\bfx=\frac{\bfs_\bfx\times\bfs_{\bfx+\bfe_1}}{\left|\bfs_\bfx\times\bfs_{\bfx+\bfe_1}\right|}.
    \label{kiral3}
\end{equation}
The additional discrete order parameter in the $K=2$ case is defined as
\begin{equation}
    \sigma_\bfx=\bfk_{\bfx,1}\cdot\bfk_{\bfx,2},
\end{equation}
where $\bfk_{\bfx,\mu}$ is defines similarly to (\ref{kiral3}) in the lattice direction $\bfe_\mu$.

\section{Results}
\label{sect3}

\subsection{$N=2$, $K=1$}

\begin{figure}[t]
\center
\includegraphics[width=\linewidth]{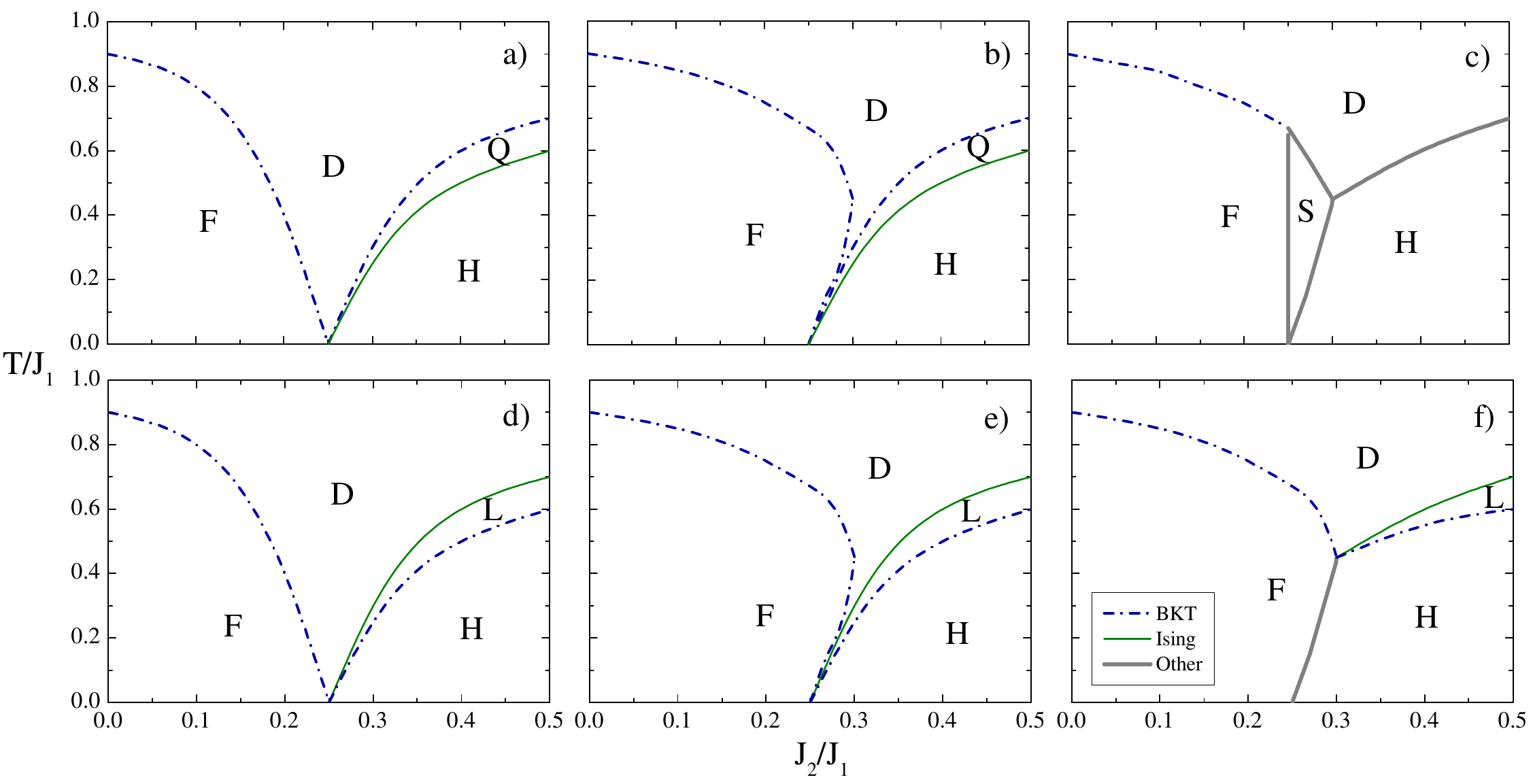}
\caption{\label{fig1}Possible phase diagrams of the $N=2$, $K=1$ helimagnet discussed in the previous studies: a) \cite{Garel80,Cinti11}; b) \cite{Okwamoto84}; c) \cite{Saslow98}; d) \cite{Kolezhuk00, Sorokin12}; e)\cite{Pokrovsky14,Dimitrova14}. Phases are denoted by the letters: F --- ferromagnetic with quasi-long-range order, H --- helicoidal, P --- paramagnetic (disordered), L --- chiral spin liquid, Q --- quasi-long-range planar order, S --- smectic-like phase. }
\end{figure}%
In contrast to the rest cases, studies of the $N=2$, $K=1$ case have a long history, which can be considered more or less completed in 2014. Already in the earliest work \cite{Garel80}, the critical behavior in the two-dimensional XY helimagnet has been associated with the spectrum of topological defects. Arguing that below an Ising transition temperature $T<T_\mathrm{dw}$ domain walls have a non-zero tension, the authors \cite{Garel80} have concluded that vortices have the linearly growing interaction (confinement) additional to the usual logarithmical one, so a BKT transition occurs at temperature above an Ising one $T_\mathrm{v}>T_\mathrm{dw}$. The proposed phase diagram is shown in fig\ref{fig1}a. Okwamoto have noted \cite{Okwamoto84} that in the vicinity of the Lifshitz point one observe the reentrant phase transition to the collinear quasi-long-range ordered phase (fig.\ref{fig1}b).

Of cause, a XY helimagnet is not only system with the $\mathbb{Z}_2\otimes SO(2)$ order parameter space. A the same time, a XY triangular antiferromagnet, the fully frustrated XY model \cite{Villain77} (FFXY) describing a superconducting array of Josephson junctions under an external transverse magnetic field, the Ising-XY model and etc. are also discussed and even more intensively (see \cite{Korshunov06} for a review). Importantly, it has been found in the studies of such systems that a crucial role in determining the critical and thermal properties of a system is played by topological defects such as $SO(2)$ vortices, $\mathbb{Z}_2$ domain walls, as well as kinks propagating on these walls. Some kinds of kinks produce an additional vorticity and behave as fractional vortices. The logarithmical interaction of kinks is weaker than the interaction of the conventional vortices and leads to a phase transition on a domain wall at $T_\mathrm{fv}<T_\mathrm{v}$. At $T>T_\mathrm{fv}$, the domain wall turns opaque for the correlations of a phase parameter describing spin orientation. As a consequence, on approaching the continuous Ising-like transition, the quasi-long-range $SO(2)$ order has to break down, and a BKT-transition has to occurs at $T_\mathrm{v} < T_\mathrm{dw}$ \cite{Korshunov02}. These arguments can be expected to be valid for any system where some kinds of kinks are fractional vortices. In such systems, a BKT-transition has to occur at temperature below an Ising transition temperature, or both transitions can occur at the same temperature as a first-order transition. The mutual influence of different topological defects and their impact on a possible sequence of phase transitions were considered for the FFXY model and for a triangular antiferromagnet \cite{Halsey85,Korshunov86,Korshunov86-2}, as well as for a special model of the helimagnet \cite{Villain91,Uimin94}.

In contrast to the FFXY model and a triangular antiferromagnet, a frustrated helimagnet is anisotropic, because it has the emphasized direction (on a lattice) corresponding to the helix. Surely, domain walls directed along the helix and across the helix have a different tension. In particular, near the Lifshitz point, a domain wall directed across the helix is very light, so a domain with the chirality of the opposite sign also directed across the helix can appear practically without an energy loss. This fact has allowed authors in ref. \cite{Saslow98} to suppose that near the Lifshitz point a helimagnet has the smectic-like phase (fig.\ref{fig1}c). But more rigorous approaches \cite{Okwamoto84,Pokrovsky14,Dimitrova14} show that this phase corresponds to the collinear quasi-long-range order (figs.\ref{fig1}e or \ref{fig1}f). Nevertheless, very light domain walls bring troubles in simulations: a thermalization from a random spin configuration almost always leads to a configuration with a domain structure. Moreover, domains may appear and disappear during simulations that leads to not correct results. For example, the appearance of a domain structure has been assumed as the chiral transition in \cite{Cinti11}.

In very interesting works \cite{Pokrovsky14,Dimitrova14} the vicinity of the Lifshitz point has been studied. In addition to the reentrant phase transition to the collinear quasi-long-range ordered phase, the authors have found that the anisotropy of the system leads to a non-Ising character of the chiral transition modified by a long-range interaction. In other words, the chiral transition falls to the universality class of the two-dimensional Ising model with strong long-range dipolar interactions \cite{Dimitrova14}. But in our opinion, this conclusion is the consequence of the chosen approximation, when a domain wall along the helix direction is infinitely heavy. As long as the domain wall tension remains finite and is influenced by vortices and fractional vortices, such a anisotropy does not refute the arguments \cite{Korshunov02} discussed above. We expect that one observes a crossover behavior with approaching the chiral transition temperature:  initially,  on a small scale, the critical behavior is of the Ising model with dipolar interactions, but in the thermodynamical limit, the behavior becomes similar to the usual Ising model. Unfortunately, the BKT, chiral and reverse (reentrant) transitions occur at very close temperatures, so we cannot investigate the critical behavior in our simulations.

\begin{figure}[t]
\center
\includegraphics[scale=0.4]{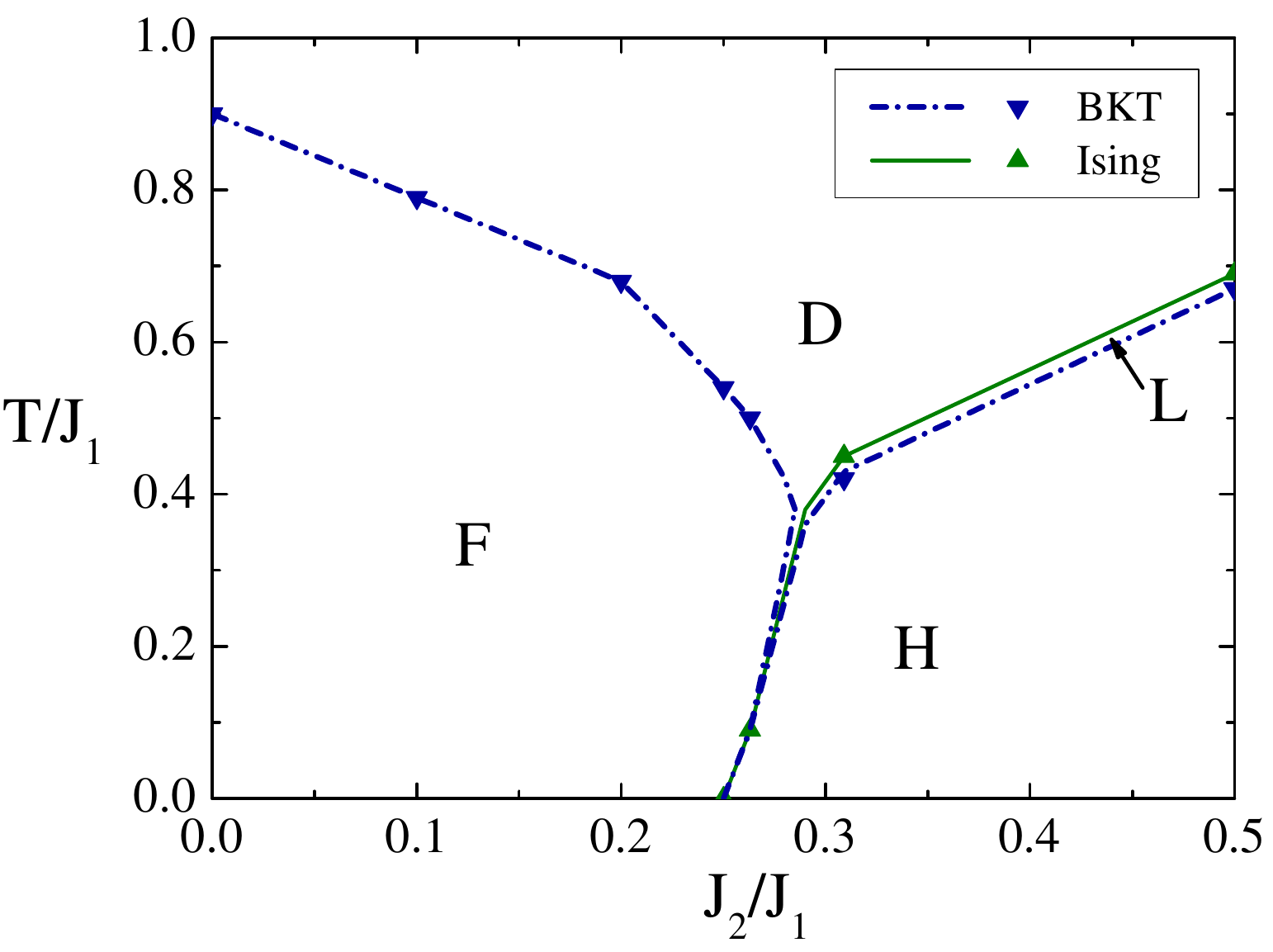}
\caption{\label{fig2}Phase diagrams of the $N=2$, $K=1$ helimagnet. The phases are marked the same as in fig.\ref{fig1}.}
\end{figure}%
In our previous study \cite{Sorokin12}, we have not considered the vicinity of the Lifshitz point and have found the simplistic phase diagram (fig.\ref{fig1}d). In the current study, we extend the results and find the diagram (fig.\ref{fig2}) consistent with  \cite{Pokrovsky14,Dimitrova14}.

\begin{figure}[t]
\center
\includegraphics[scale=0.4]{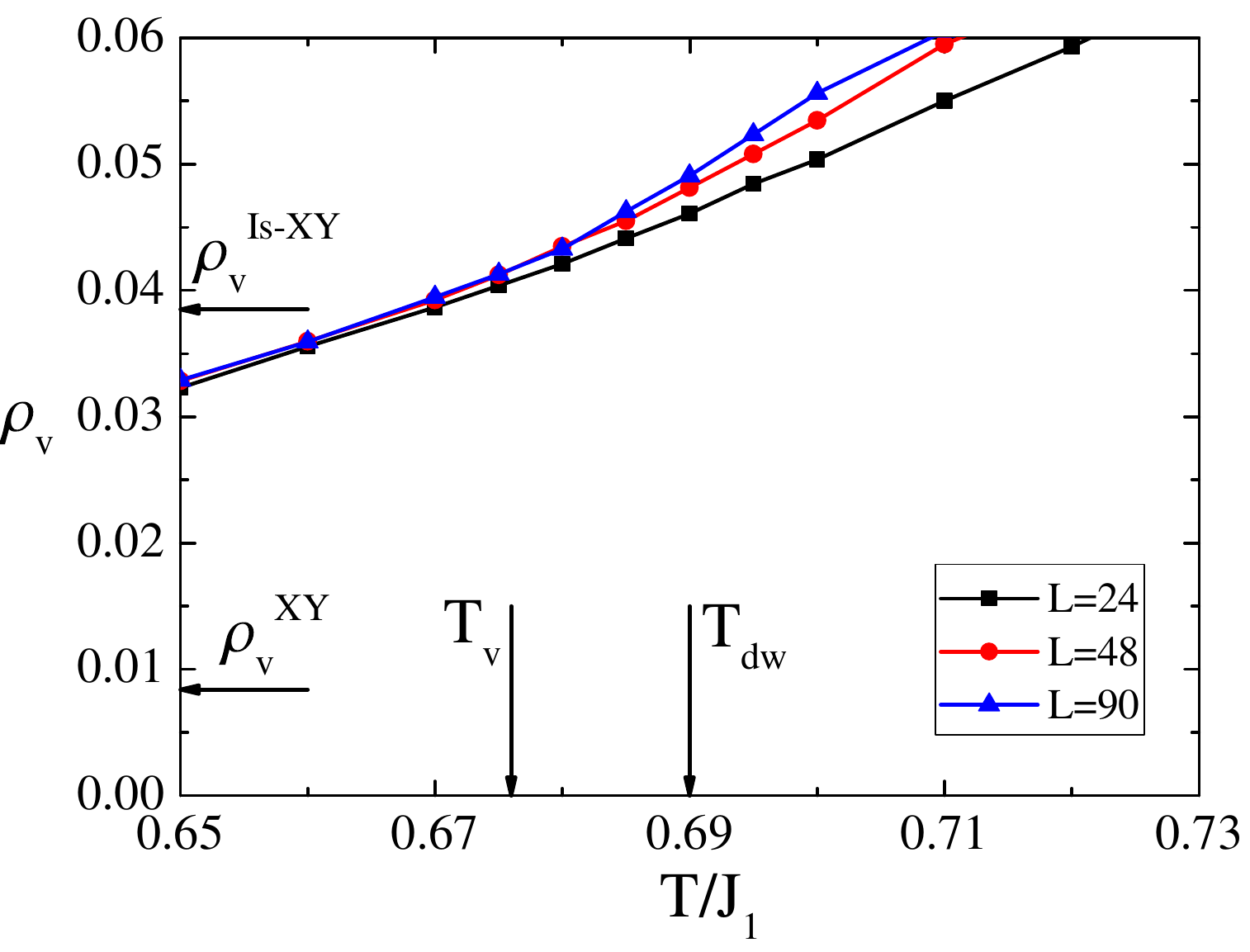}
\caption{\label{fig3}Vortex density for the case $J_2/J_1=0.5$.}
\end{figure}%
We also investigate the critical properties of topological defects, namely domain walls and vortices. The density of vortices is
\begin{equation}
    \tilde\rho_\mathrm{v}=\frac{1}{2\pi L^2}\sum_{\bfx}\sum_{\square_\bfx}\varphi_{ij},
    \quad \rho_\mathrm{v}=\langle\tilde\rho_\mathrm{v}\rangle,
\end{equation}
where $x$ runs over primitive cells of a lattice, and $\square_\bfx$ means coming over a cell and summing differences of spin phases $\varphi_{ij}=\varphi_i-\varphi_j\in(-\pi,\pi]$. For the case $J_2/J_1=0.5$, the thermal dependence of the vortex density is shown in fig.\ref{fig3}. At the BKT transition point $T_\mathrm{v}=0.676(2)$, the critical value of the vortex density is
\begin{equation}
    \rho_\mathrm{v}=0.042(3).
\end{equation}
This value is much larger than the value of the pure $O(2)$ model \cite{Sorokin18}: $\rho_\mathrm{v}\approx0.0084$, but is in consistent with the value of the vortex density at the BKT transition point of the Ising-XY model \cite{Sorokin18-2}: $\rho_\mathrm{v}\approx0.037$. As it has been discussed in \cite{Granato91,Lee91,Kosterlitz91}, the special case of the Ising-XY model corresponds to the FFXY model and a triangular antiferromagnet as well as a XY helimagnet too.

\begin{figure}[t]
\center
a)
\includegraphics[width=0.2\textwidth]{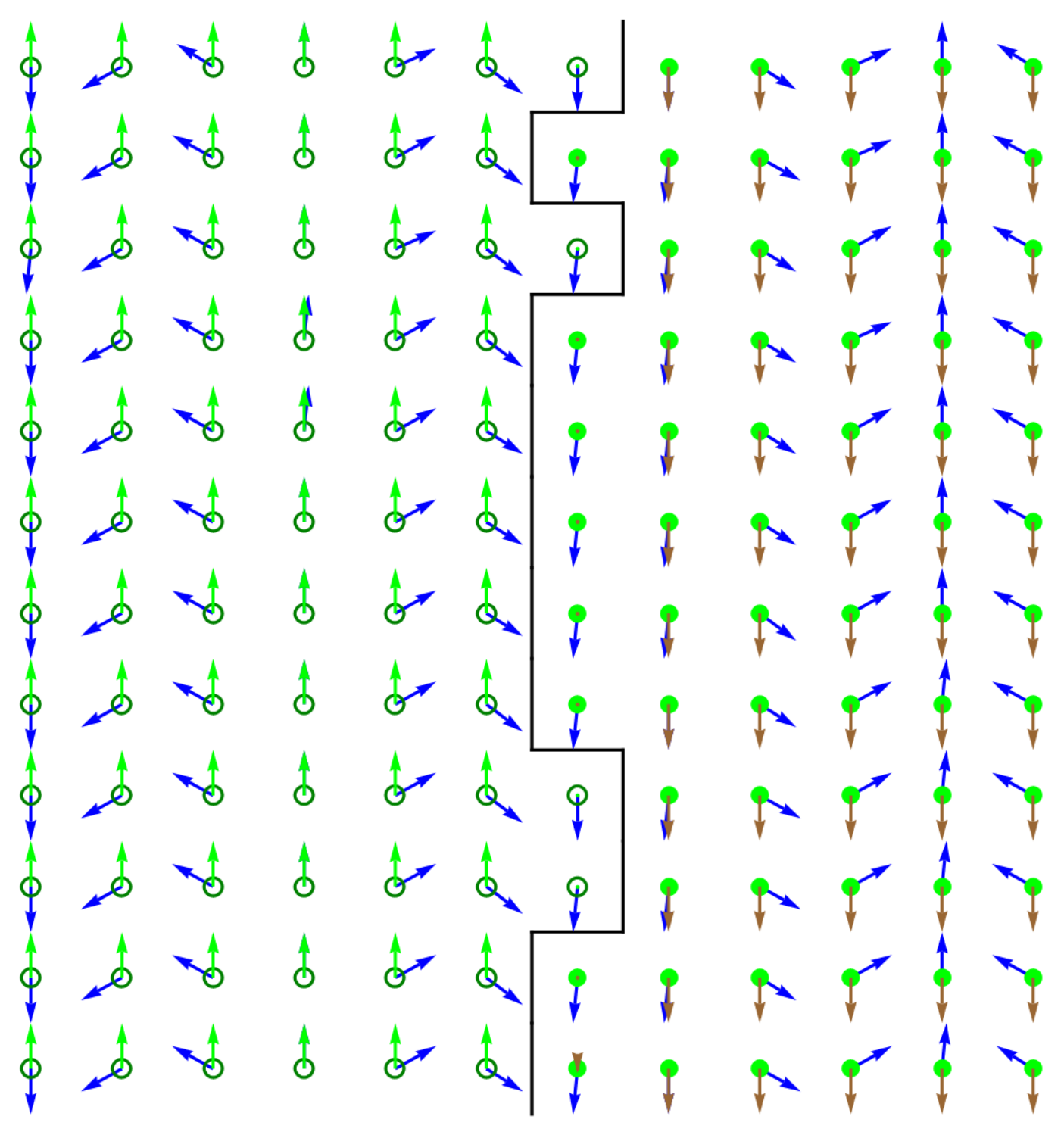}
b)
\includegraphics[width=0.2\textwidth]{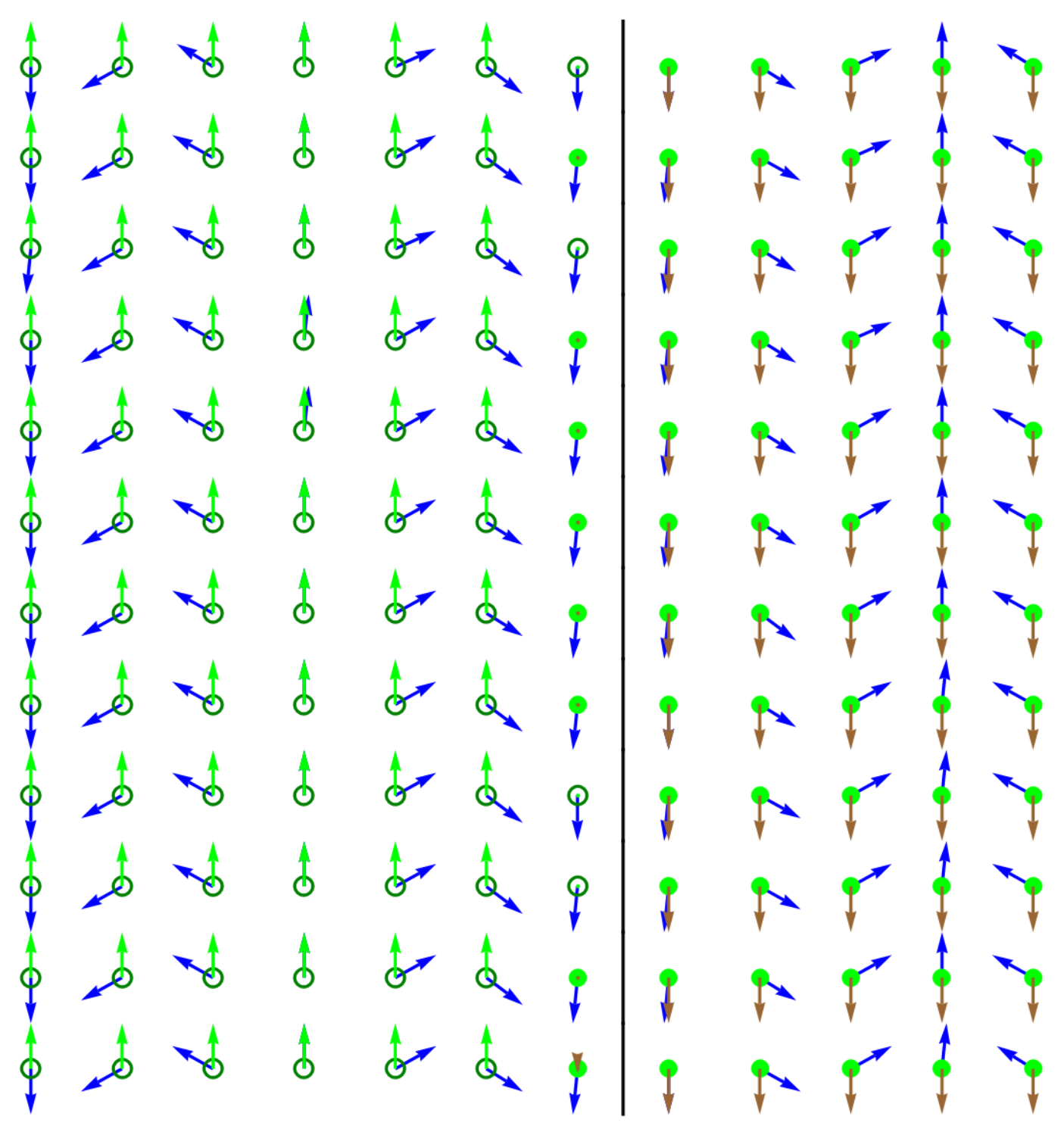}
\caption{\label{fig4}Single domain wall at $T/J_1=0.001$: a) naive determination of a wall position; b) improved determination of a wall position.}
\end{figure}%
Since, we have the local definition of the chiral (Ising-like) order parameter $k_\bfx=\frac{1}{\sin\theta_0}\sin(\varphi_\bfx-\varphi_{\bfx+\bfe_\mu})$, $\sigma_\bfx=\frac{\bfk_x}{|\bfk_x|}$, the domain wall density can be naively defined as
\begin{equation}
    \tilde\rho_\mathrm{dw}^\mu=\frac{1}{4L^2}\sum_\bfx(1-\sigma_\bfx\sigma_{\bfx+\bfe_\mu}),
    \quad \rho_\mathrm{dw}=\sum_\mu\langle\tilde\rho_\mathrm{dw}^\mu\rangle.
    \label{ro-dw-ising}
\end{equation}
But this definition is not correct. In contrast to the pure Ising model, the order parameter $k_\bfx$ is not normalized. To understand troubles related to this definition, lets consider a single domain wall directed across the helix. At the equilibrium and zero temperature, the spin configuration corresponding to this wall has bonds with parallel spins, so we see the line of bonds with zero chirality. At finite temperature, this line is perturbed, and even very small fluctuations change the chirality along the line from zero value to some non-zero one (see fig.\ref{fig4}a). Korshunov has proposed \cite{Korshunov} the alternative determination of a wall position: a wall passes through a bond with the minimal value of the chirality. In details, lets $k_\bfx k_{\bfx+\bfe_1}<0$ and $|k_\bfx|<| k_{\bfx+\bfe_1}|$ then a domain wall passes through the bond with $k_\bfx$. This rule is not correct too, because it is possible that two walls pass through one bond. For Ising-like ($\mathbb{Z}_2$) walls, it is equivalent to no wall passes. Finally, we define:
\begin{equation}
    \sigma_\bfx=\left\{
    \begin{array}{cc}
      -\frac{\bfk_x}{|\bfk_x|} & \begin{array}{c}
                                   ((k_\bfx k_{\bfx-\bfe}\leq 0)\wedge(|k_\bfx|<|k_{\bfx-\bfe}|))\wedge \\
                                   \wedge((k_\bfx k_{\bfx+\bfe}>0)\vee(|k_\bfx|>|k_{\bfx+\bfe}|)) 
                                 \end{array}
       \\
      \frac{\bfk_x}{|\bfk_x|} & \mathrm{other\ cases} 
    \end{array}
    \right..
\end{equation}
And now the domain wall density can be defined as (\ref{ro-dw-ising}). 

\begin{figure}[t]
\center
a)
\includegraphics[width=0.2\textwidth]{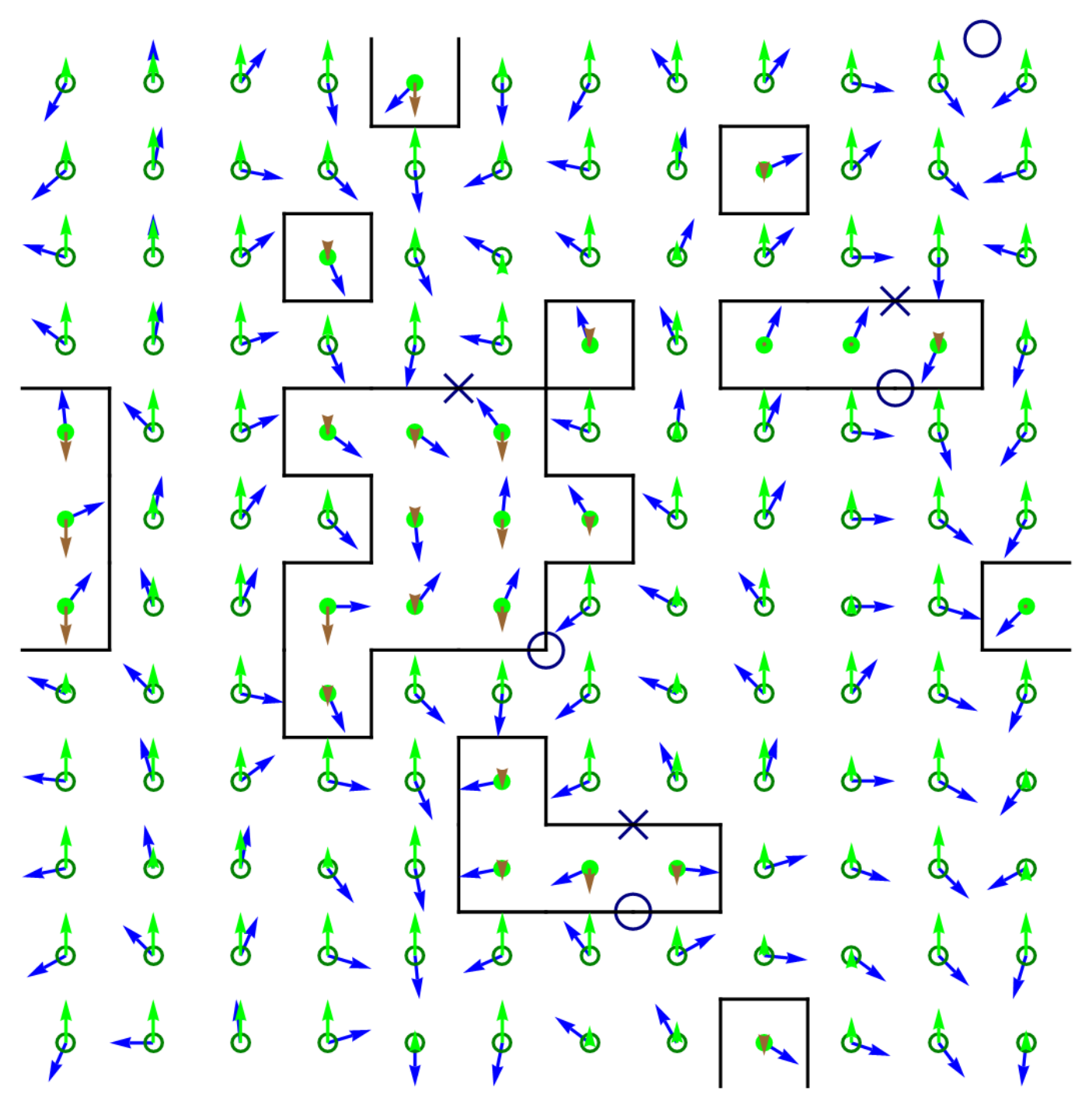}
b)
\includegraphics[width=0.2\textwidth]{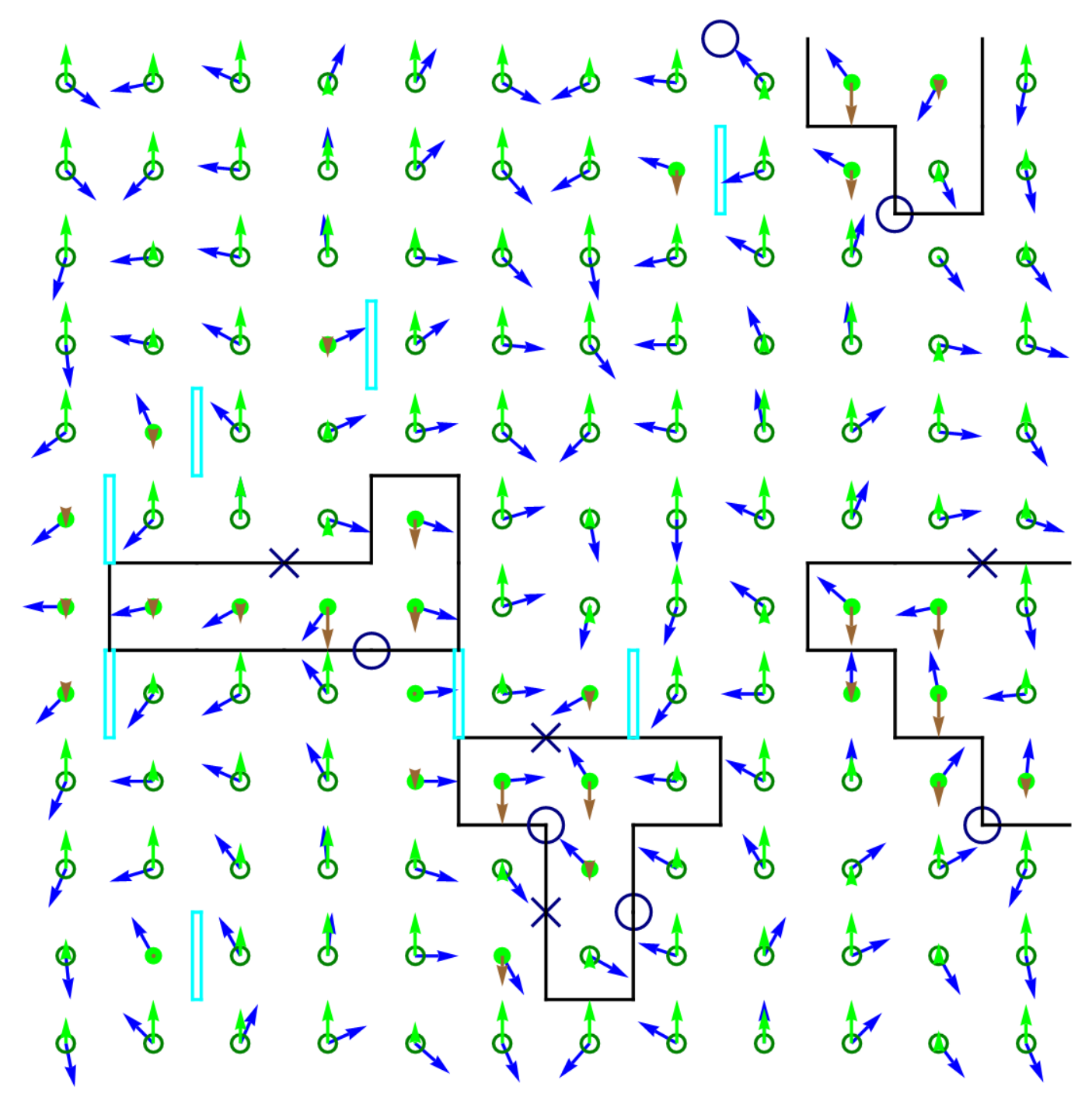}
\caption{\label{fig5}Shot of a simulation at $T/J_1=0.65$: a) naive determination of a wall position; b) improved determination of a wall position.}
\end{figure}%
\begin{figure}[t]
\center
\includegraphics[scale=0.4]{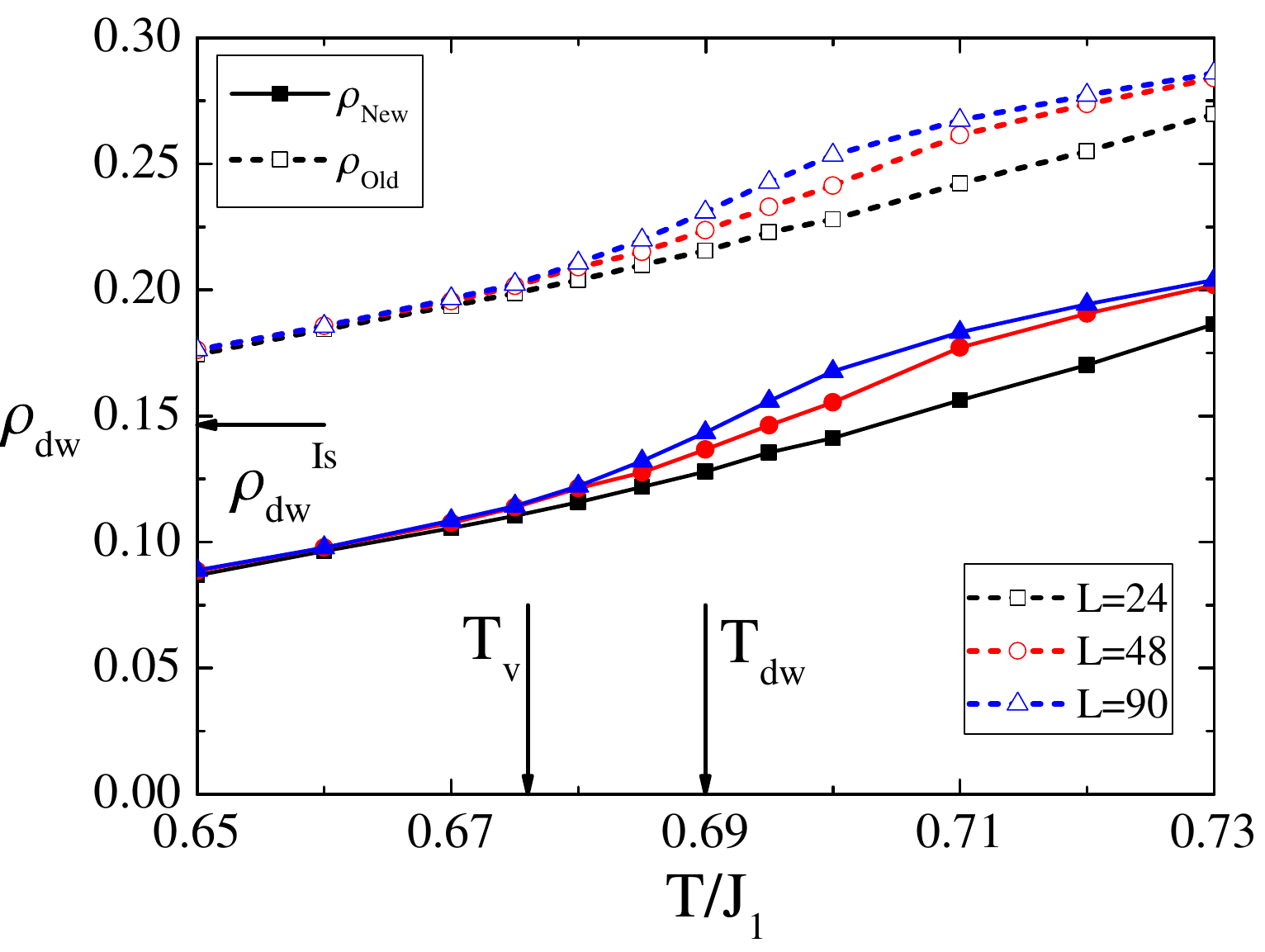}
\caption{\label{fig6}Domain wall density for the case $J_2/J_1=0.5$.}
\end{figure}%
This not elegant definition leads to the elimination of non-physical modes of a domain wall (see fig.\ref{fig4}b). Also, it excludes bonds with zero chirality, which may be considered as a non-zero thickness of a wall, and excludes loop with zero square (fig.\ref{fig5}). But the main advantage of improved definition is the critical value of the wall density at the chiral transition temperature $T=T_\mathrm{dw}$. Fig\ref{fig6} shows that the naive determination of a wall position predicts a overestimated density value, while the improved determination gives the value
\begin{equation}
    \rho_\mathrm{dw}=0.149(4),
\end{equation}
that is in agreement with the results for the Ising model \cite{Sorokin18} as well as for the Ising-XY model \cite{Sorokin18-2}.

\begin{figure}[t]
\center
\includegraphics[scale=0.4]{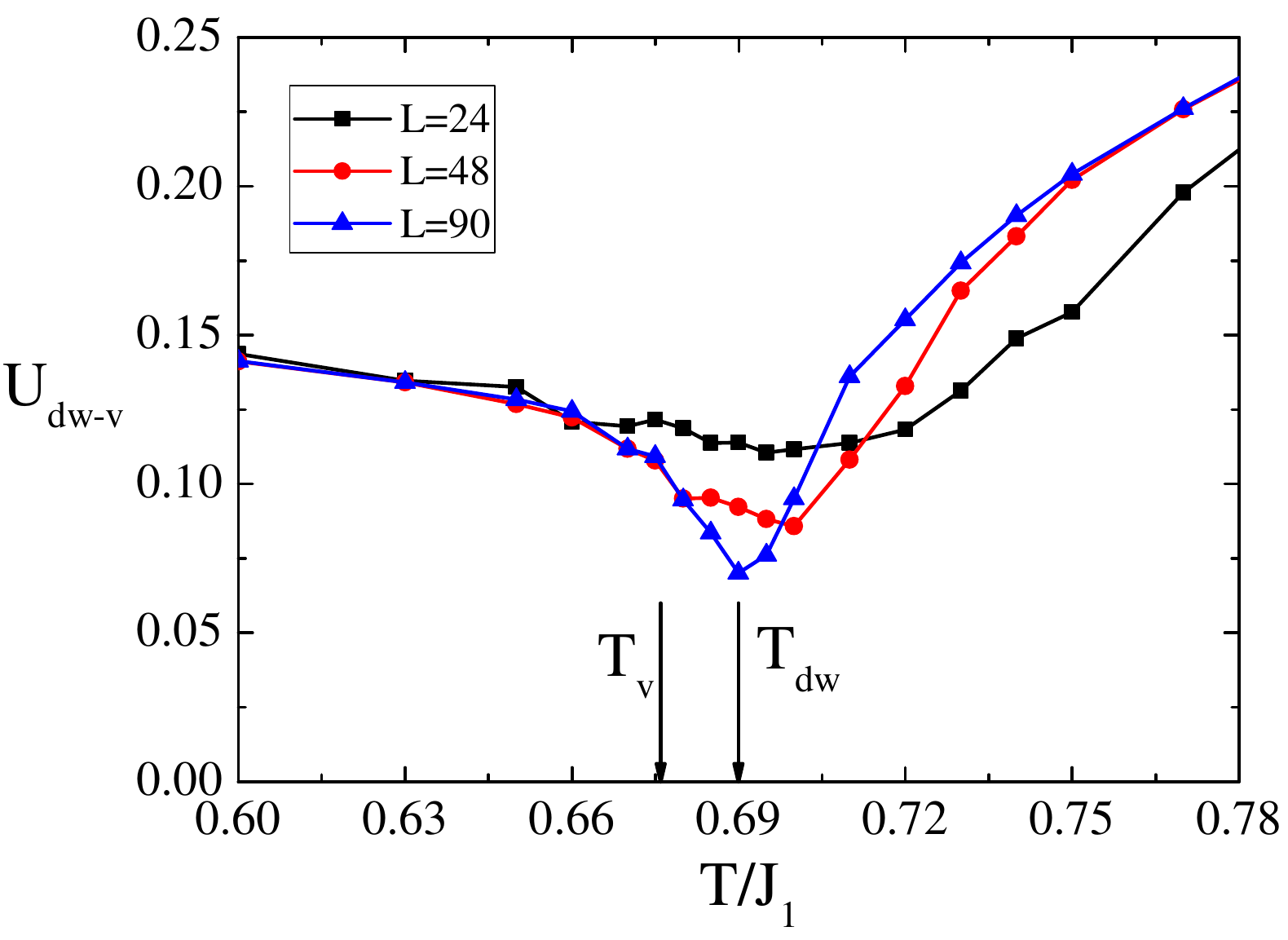}
\caption{\label{fig7}Wall-vortex correlator for the case $J_2/J_1=0.5$.}
\end{figure}%
Following the paper \cite{Sorokin18-2}, we consider the wall-vortex correlator (the Pearson correlation coefficient)
\begin{equation}
    U_{\mathrm{dw-v}}\equiv
    \frac{\langle\rho_\mathrm{dw}\rho_\mathrm{v}\rangle-\langle\rho_\mathrm{dw}\rangle\langle\rho_\mathrm{v}\rangle}{\sqrt{\chi_\mathrm{dw}\chi_\mathrm{v}}},
    \label{td-td-corr}
\end{equation}
where $\chi_\mathrm{td}$ is the topological susceptibility
\begin{equation}
\chi_\mathrm{td}=L^2\left(\langle\rho_\mathrm{td}^2\rangle-\langle\rho_\mathrm{td}\rangle^2\right).
\label{chi}
\end{equation}
The cumulant $U$ has the meaning of the effective coupling constant of local defect-defect interactions. As long as the topological defect densities are jointly normally distributed quantities, the vanishing of the cumulant $U$ means that topological defects have no local interactions, both linear or non-linear. One can see in fig.\ref{fig7} that the correlator tends to zero in the thermodynamical limit at the chiral transition point. This explains why vortices do not affect the wall density at the critical point: at $T=T_\mathrm{dw}$ wall-vortex interactions become insignificant and vanish with $L\to\infty$.

\subsection{$N=2$, $K=2$}

\begin{figure}[t]
\center
\includegraphics[scale=0.4]{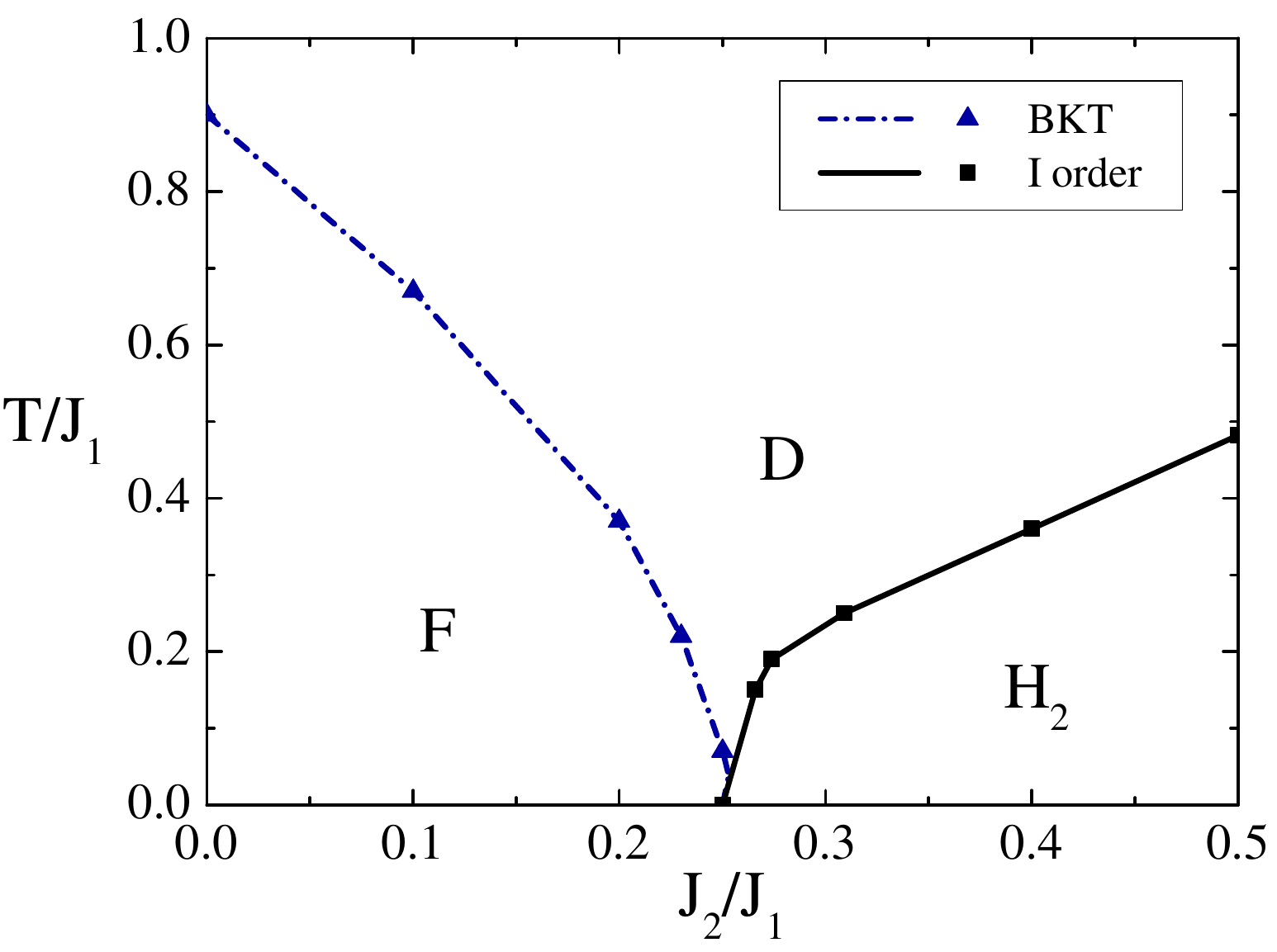}
\caption{\label{fig8}Phase diagrams of the $N=2$, $K=2$ helimagnet. The phases are marked the same as in fig.\ref{fig1}.}
\end{figure}%
The case of XY helimagnets with two chiral order parameter has been considered in \cite{Sorokin12-2}. The isotropic subcase $J_3^1=J_3^2$ is considered in detail in the current work. In contrast to the anisotropic subcase $J_3^1\neq J_3^2$, where we have found the possibility of separated in temperature phase transition and chiral spin liquid phase existence \cite{Sorokin12-2}, here we find the single first-order transition for all values of the exchange constant ratio $J_2/J_1>0.25$ (fig.\ref{fig8}).

\subsection{$N=3$, $K=1$}
\begin{figure}[t]
\center
\includegraphics[scale=0.4]{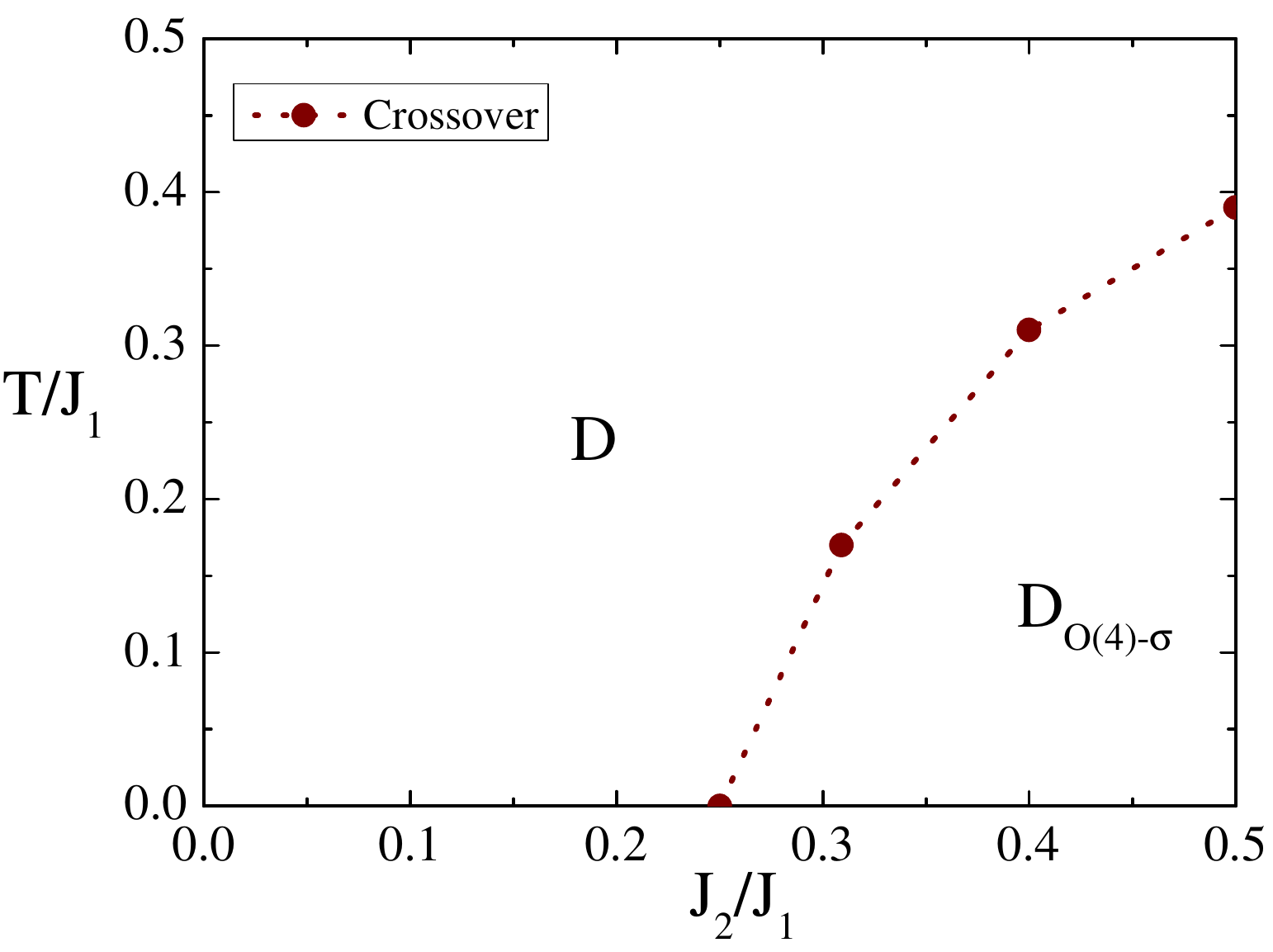}
\caption{\label{fig9}Phase diagrams of the $N=3$, $K=1$ helimagnet. Letter $D$ marks the disordered phase with a high-temperature behavior, $D_{O(4)-\sigma}$ is the phase with the behavior predicted by the $O(4)$ $\sigma$-model.}
\end{figure}%
\begin{figure}[t]
\center
\includegraphics[scale=0.4]{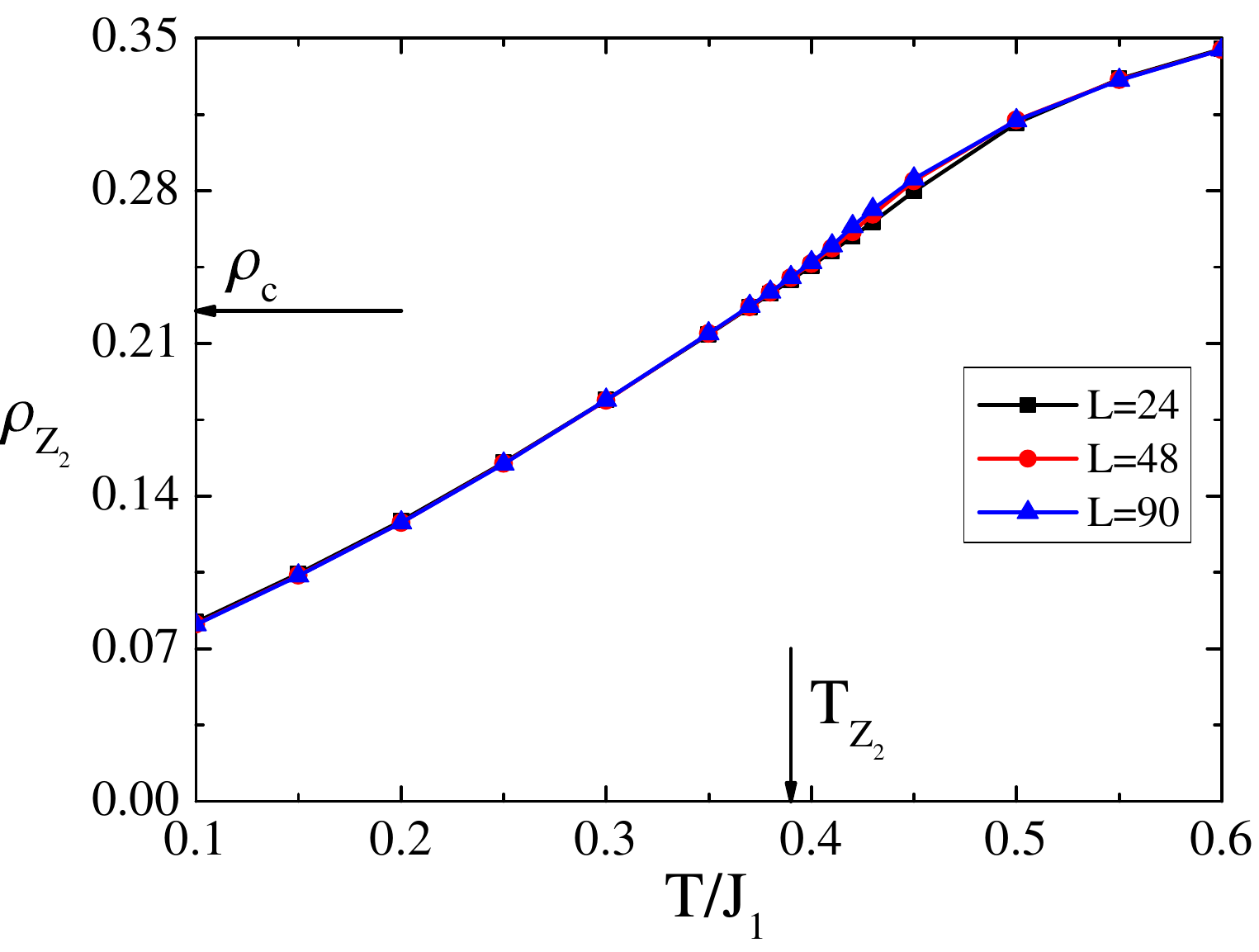}
\caption{\label{fig10}Vortex density for the case $J_2/J_1=0.5$.}
\end{figure}%
A simple ($K=1$) helimagnet with isotropic spins ($N=3$) has not been previously investigated. However, it is not only system with so-called $\mathbb{Z}_2$ vortices in the spectrum of topological excitations. Such vortices arise in systems with the order parameter space $G/H$, the fundamental group of which is $\pi_1(G/H)=\mathbb{Z}_2$. For example, such a topology corresponds systems with $G/H=SO(N)$, $N\geq3$, or $G/H=\mathbb{R}P^{N-1}$, $N\geq3$. The first example is the order parameter space of the $V_{N,N-1}$ Stiefel model \cite{Zumbach93}, including frustrated magnets. The case $G/H=SO(3)$ has been first considered in \cite{Kawamura84} for a $N=3$ triangular antiferromagnet. The case $G/H=\mathbb{R}P^2$ corresponding to nematics has been studied in \cite{Duane81}. Authors of earliest works have expected that $\mathbb{Z}_2$ vortices lead to a BKT-like transition. But in the more recent works \cite{Azaria01,Hasenbusch96,Niedermayer96,Sinner14}, it has been shown that there is no long-range or quasi-long-range orders at finite temperature. Nevertheless, one observes a crossover between the $O(4)$ $\sigma$-model behavior and the high-temperature behavior induced by $\mathbb{Z}_2$ vortices. Recently, we have also considered the system with $G/H=SO(3)$, namely the $V_{3,2}$ Stiefel model \cite{Sorokin17,Sorokin18}.

As expected, the temperature behavior of a helimagnet demonstrates an explicit crossover at finite temperature (fig.\ref{fig9}). We also find that the $\mathbb{Z}_2$ vortex density defined in the same way as in \cite{Sorokin17,Sorokin18,Sorokin18-2} has the critical value (fig.\ref{fig10})
\begin{equation}
    \rho_{\mathbb{Z}_2}=0.237(15),
\end{equation}
that is in agreement with the result for the $V_{3,2}$ Stiefel model \cite{Sorokin18}.

\subsection{$N=3$, $K=2$}

\begin{figure}[t]
\center
\includegraphics[scale=0.4]{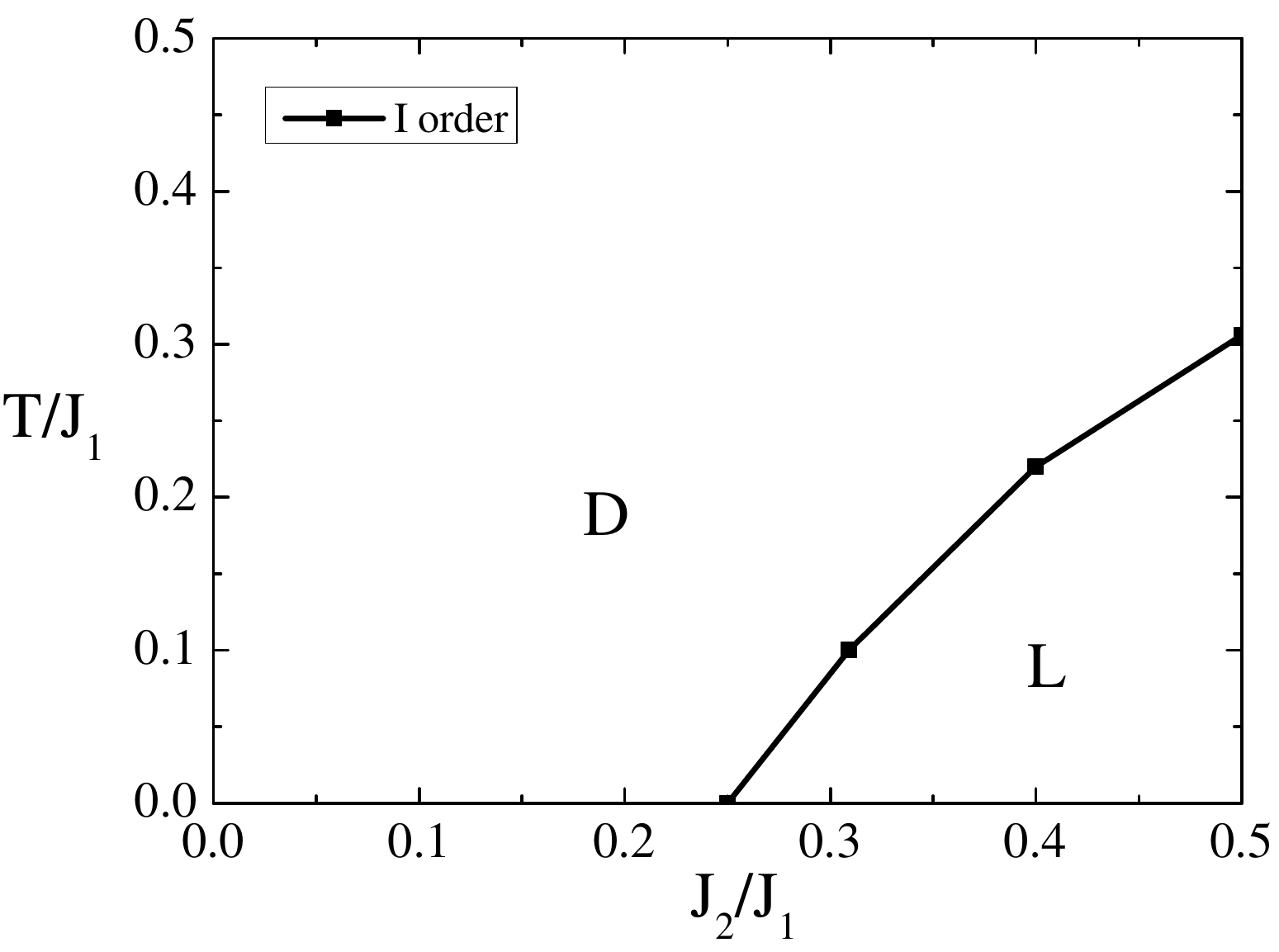}
\caption{\label{fig11}Phase diagrams of the $N=3$, $K=2$ helimagnet. The phases are marked the same as in fig.\ref{fig1}.}
\end{figure}%
The case $N=3$, $K=2$ of the model (\ref{model}) have been intensively studied for quantum spins. It is known as J$_1$-J$_2$-J$_3$ model. But classical spins have been considered too in the work \cite{Sachdev04}. The authors have found the single second-order phase transition of the Ising type. However, we have investigated the case $J_2/J_1=0.5$ in \cite{Sorokin17} and have found that the transition is of first order. Similar behavior observes in the $V_{3,3}$ Stiefel model.

One may expect that the type of the transition changes near the Lifshitz point. But our data confirm the first order of the transition for other values of the exchange constant ratio (fig.\ref{fig11}). It is important that the first order of the transition is induced by $\mathbb{Z}_2$ vortices. We find that the transition in the discrete order parameter as well as the crossover in the thermal behavior of the $SO(3)$ order parameter occur at the same temperature.

Note also that the phase below the transition temperature corresponds to the long-range order in the discrete order parameter without long-range or quasi-long-range orders in the continuous parameter. Similar to the case $N=2$, $K=1$, this phase is a classic spin liquid.

\section{Conclusion}

The critical behavior of two-dimensional frustrated helimagnets turns to be rather expected consistent with results for other systems with similar structure of the order parameter space. The most important results is the critical values of the topological defect densities in the $K=1$ cases both $N=1$ or 2. In ref. \cite{Sorokin18}, we have discussed several possibilities which may refute the hypothesis that a topological defect density has an universal value at the temperature of corresponding transition. The verification of this hypothesis is especially important for Ising-like transitions, where the critical properties of domain walls are determined by the conformal symmetry. One of such possibilities is the case when an interaction between vortices and domain walls determines a consequence of phase transition like in the Ising-XY model (see \cite{Sorokin18-2} for details, where this possibility has been excluded). Another one is the case when an order parameter is not normalized, and topological defects (e.g., domain walls in a XY helimagnet) have (may be, non-physical) internal degrees of freedom. In this paper, we exclude this possibility too. An improved definition of topological defects excluding internal modes leads the universal value, at least as long as a transition remain of second order.

\medskip

This work is supported by the RFBR grant No 16-32-60143.

This work have been started in a form of very useful discussions with Sergey Korshunov, but to our regret, it have been completed after his untimely death in 2016.

\end{document}